\pdfoutput=1
\RequirePackage[T1]{fontenc}
\RequirePackage{fix-cm}
\RequirePackage{lmodern}
\documentclass[12pt,pdftex]{article}

\usepackage[margin=15truemm]{geometry}
\usepackage{color}

\newcommand{\ket}[1]{|#1 \rangle}
\newcommand{\bra}[1]{\langle #1 |}
\newtheorem{example}{Example}

\newtheorem{lemma}[example]{Lemma}

\newtheorem{definition}[example]{Definition}
\newtheorem{remark}[example]{Remark}

\begin{document}
\date{21 April 2021}
\title{Constructions of $\ell$-Adic $t$-Deletion-Correcting Quantum Codes\thanks{%
  This paper is partially supported by Japan Society of Promotion of Science
  KAKENHI 18H01435.}}
\author{Ryutaroh Matsumoto\thanks{%
  Department of Information and Communications Engineering,
  Tokyo Institute of Technology, 152-8550 Japan, and
  Department of Mathematical Sciences, Aalborg University, Aalborg, Denmark
  (email: ryutaroh@ict.e.titech.ac.jp)} \and
Manabu Hagiwara\thanks{%
Department of Mathematics and Informatics, Graduate School of Science, Chiba University,
  1-33 Yayoi-cho, Inage-ku, Chiba City, Chiba Pref., 263-0022 Japan
  (email: hagiwara@math.s.chiba-u.ac.jp)}}
\maketitle
\begin{abstract}
We propose two systematic constructions of deletion-correcting
codes for protecting quantum information.
The first one works with qudits of any dimension, but only one deletion is
corrected and the constructed codes are asymptotically bad.
The second one corrects multiple deletions and can construct
asymptotically good codes.
The second one also allows conversion of stabilizer-based quantum codes
to deletion-correcting codes, and entanglement assistance.
\end{abstract}


\section{Introduction}
In  the context of conventional (classical) error correction,
deletion correction, which was introduced by Levenshtein in 1966
\cite{levenshtein66}, has attracted much attention recently
(see, for example, \cite{sima21} and the references therein).
In the correction of erasures,
the receiver is aware of positions of erasures \cite{bennett97,grassl97,macwilliams77}.
In contrast to this, the receiver is unaware of positions of deletions,
which adds extra difficulty to correction of deletions and code constructions
suitable for deletion correction.
Partly due to the combined difficulties of deletion correction and
quantum error correction, the study of quantum deletion correction
has begun very recently \cite{hagiwara20b,hagiwara20a,hagiwara20c}.
Those researches provided concrete examples of quantum deletion-correcting codes.
The first systematic construction of
$1$-deletion-correcting binary quantum codes was proposed in \cite{hagiwara20b},
where $((2^{k+2}-4, k))_2$ codes were constructed for any positive integer $k$.
  Very recently, the first systematic constructions of
  $t$-deletion-correcting binary quantum codes were proposed \cite{hagiwara21,picode2}
  for any positive integer $t$.
  The number of codewords was two in \cite{hagiwara21,picode2}.
There are the following problems in the existing studies:
(1) There is no systematic construction for nonbinary quantum codes correcting
more than $1$ deletions.
(2) Existing studies of quantum error correction cannot be reused
in an obvious manner.

In this paper, we tackle these problems by proposing two systematic
constructions of nonbinary quantum codes.
The first one is based on the method of type in the information theory \cite{csiszarbook2}.
The constructed codes belong to the class of permutation-invariant quantum codes
\cite{picode,picode2}. It can construct quantum codes for qudits of arbitrary dimension,
but the codes can correct only $1$ deletion and asymptotically bad.
The second construction converts quantum erasure-correcting codes
to deletion-correcting ones. The construction is asymptotically good,
and can correct as many deletions as the number of correctable erasures
of the underlying quantum codes.
But the second construction has severe limitations on the dimension of qudits.
For example, the second construction cannot construct binary or ternary quantum codes.

This paper is organized as follows: Section \ref{sec2}
introduces necessary notations and concepts.
Section \ref{sec3} proposes the first construction.
Section \ref{sec4} proposes the second construction.
Section \ref{sec5} concludes the paper.

\section{Preliminaries}\label{sec2}
Let $\mathbf{Z}_\ell = \{0$, $1$, \ldots, $\ell-1\}$.
A type $P$ \cite{csiszarbook2}
of length $n$ on the alphabet $\mathbf{Z}_\ell$
is a probability distribution on $\mathbf{Z}_\ell$
such that each probability in $P$ is of the form $m/n$, where $m$ is an integer.
The alphabet is fixed to $\mathbf{Z}_\ell$
when we consider types.
For $\vec{x} = (x_1$, \ldots, $x_n) \in \mathbf{Z}_\ell^n$,
the type $P_{\vec{x}}$ of $\vec{x}$ is
the probability distribution $P_{\vec{x}}(a) = \sharp \{ i \mid x_i = a \} / n$,
where $\sharp$ denotes the number of elements in a set.
For a type $P$ of length $n$,
$T(P)$ denotes the set of all sequences with type $P$, that is,
\[
T(P) = \{ \vec{x} \in \mathbf{Z}_\ell^n \mid P_{\vec{x}} = P \}.
\]
For types $P_1$ and $P_2$, we define $P_1 \sim P_2$
if there exists a permutation $\sigma$ on $\ell$ numbers in a type  such that
$\sigma(P_1) = P_2$.
For example, when $P_1 = (1/3$, $1/6$, $1/2)$, $\sigma(P_1)$
can be $(1/6$, $1/2$, $1/3)$.
This $\sim$ is an equivalence relation, and we can consider
equivalence classes induced by $\sim$.
We denote an equivalence class represented by $P$ by $[P]$.
We define $T([P]) = \bigcup_{Q \in [P]} T(Q)$.

\begin{definition}
  For $0 \leq t \leq n-1$, we say
  a type $P_1$ of length $n-t$ to be a type of $P_2$ after $t$ deletion,
  where $P_2$ is a type of length $n$, if
  \begin{itemize}
    \item For each $a \in \mathbf{Z}_\ell$,
      $(n-t)P_1(a) \leq n P_2(a)$,
    \item and $\sum_{a \in \mathbf{Z}_\ell} \{n P_2(a) - (n-t)P_1(a)\} = t$.
  \end{itemize}
\end{definition}
We see that $P_{\vec{y}}$ is a type of $P_{\vec{x}}$ after $t$ deletion
if $\vec{y}$ is obtained by deleting $t$ components in $\vec{x}$.

\begin{definition}
  Let $S = \{ P_0$, \ldots, $P_{M-1} \} $ be a set of types of length $n$.
  We call $S$ to be suitable for $t$-deletion correction
  if for any $Q_1 \in [P_i]$ and any $Q_2 \in [P_j]$ with $Q_1 \neq Q_2$
  there does not exist a type $R$ of length $n-t$
  such that $R$ is a type of both $Q_1$ and $Q_2$ after $t$ deletion.
\end{definition}

Let $\mathcal{H}_\ell$ be the complex linear space of
dimention $\ell$.
By an $((n,M))_\ell$ quantum code we mean
an $M$-dimentional complex linear subspace $Q$ of
$\mathcal{H}_\ell^{\otimes n}$.
An $((n,M))_\ell$ code is said to be $\ell$-adic.
The information rate of $Q$ is defined to be $(\log_\ell M) / n$.
A code construction is said to be asymptotically good
if it can give a sequence of codes with which $\liminf_{n\rightarrow\infty} (\log_\ell M) / n >0$
\cite{macwilliams77}, and said to be bad otherwise.

\section{First Construction of Quantum Deletion Codes}\label{sec3}
\subsection{Construction}
With a given $S$ suitable for $t$-deletion correction,
we construct $((n,M))_\ell$ quantum code as follows:
An $M$-level quantum state
$\alpha_0 \ket{0} + \cdots + \alpha_{M-1}\ket{M-1}$
is encoded to a codeword $\ket{\varphi} \in Q$ as
\[
\sum_{k=0}^{M-1} \alpha_k \frac{1}{\sqrt{\sharp  T([P_k])}} \sum_{\vec{x} \in T([P_k])}
\ket{\vec{x}}.
\]
In the next subsection, we will prove this construction can correct $t=1$ deletion.

\subsection{Proof of $1$-Deletion Correction}
We assume $t=1$ in this subsection (see Remark \ref{rem1}).
The proof argument does not work for $t>1$.
Firstly, for any codeword $\ket{\varphi} \in Q$,
any permutation of $n$ qudits in $\ket{\varphi}$ does not
change $\ket{\varphi}$.
Our constructed codes are instances of the permutation-invariant
quantum codes \cite{picode,picode2}.
So any $t$ deletion of $\ket{\varphi}$ is the same as deleting
the first, the second, \ldots, the $t$-th qudits in
$\ket{\varphi}$.
Therefore, $t$ deletion on $\ket{\varphi} \in Q$
can be corrected by assuming $t$ erasures in
the first, the second, \ldots, the $t$-th qudits.

By using Ogawa et al.'s condition \cite[Theorem 1]{ogawa05},
we show that the code can correct one erasure at the first qudit
by computing the partial trace $\mathrm{Tr}_{\overline{\{1\}}}[\ket{\varphi}\bra{\varphi}]$
of $\ket{\varphi}\bra{\varphi}$
over the second, the third, \ldots, and the $n$-th qudits.

Let $\ket{\varphi_k} = \frac{1}{\sqrt{\sharp  T([P_k])}} \sum_{\vec{x} \in T([P_k])}\ket{\vec{x}}$.
We first compute
$\mathrm{Tr}_{\overline{\{1\}}}[\ket{\varphi_k}\bra{\varphi_k}]$.
Let $D_1$ be the deletion map from $\mathbf{Z}_\ell^n$
to $\mathbf{Z}_\ell^{n-1}$ deleting the first component.
For $\vec{x} \in \mathbf{Z}_\ell^n$, $x_i$ denotes the $i$-component.
\begin{eqnarray*}
  && \mathrm{Tr}_{\overline{\{1\}}}[\ket{\varphi_k}\bra{\varphi_k}] \\
  &=& \frac{1}{\sharp  T([P_k])} \sum_{a,b \in \mathbf{Z}_\ell}
  \ket{a}\bra{b}  \times \sharp \{ (\vec{x}, \vec{y})
  \in T([P_k])\times T([P_k]) \\
  && \quad \mid
  x_1 = a, y_1 =b, D_1(\vec{x})=D_1(\vec{y})\}.
\end{eqnarray*}
When $a=x_1 \neq b=y_1$ and $D_1(\vec{x})=D_1(\vec{y})$
we have $P_{\vec{x}} \neq P_{\vec{y}}$.
Since there does not exist a type $R$ of length $n-1$ such that
$R$ is $P_{\vec{x}}$ after 1 deletion and  also $R$ is  $P_{\vec{y}}$
after 1 deletion, for any $k$
there cannot exist $\vec{x}$, $\vec{y} \in T([P_k])$
such that $a=x_1 \neq b=y_1$ and $D_1(\vec{x})=D_1(\vec{y})$.
On the other hand, by the symmetry of the construction,
for any $a \in \mathbf{Z}_\ell$,
$\sharp \{ (\vec{x}, \vec{y})
\in T([P_k])\otimes T([P_k]) \mid
x_1 = a=y_1, D_1(\vec{x})=D_1(\vec{y})\}$ has the same size.
Therefore, we see that
\[
\rho_k = \mathrm{Tr}_{\overline{\{1\}}}[\ket{\varphi_k}\bra{\varphi_k}]
= \frac{1}{\ell} \sum_{a\in \mathbf{Z}_\ell} \ket{a}\bra{a}.
\]

On the other hand,
by the construction,
for $k_1 \neq k_2$, $\vec{x} \in T([P_{k_1}])$,
$\vec{y} \in T([P_{k_2}])$, $D_1 (\vec{x})$ is always different from
$D_1 (\vec{y})$, which implies
\begin{equation}
\mathrm{Tr}_{\overline{\{1\}}}[\ket{\varphi}\bra{\varphi}]
= \sum_{k=0}^{M-1} |\alpha_k|^2 \rho_k = I_{\ell \times \ell} / \ell. \label{eq1}
\end{equation}
By \cite[Theorem 1]{ogawa05}, this implies that the constructed code
can correct one erasure at the first qudit,
which in turn implies one deletion correction by the symmetry of codewords
with respect to permutations. \rule{1ex}{1ex}

\begin{remark}\label{rem1}
When $t>1$, Eq.\ (\ref{eq1}) sometimes depends on the encoded quantum information,
and one cannot apply \cite[Theorem 1]{ogawa05}.
Since the number of types is polynomial in $n$
\cite{csiszarbook2}, the proposed construction is asymptotically bad.
\end{remark}

\subsection{Examples}
\subsubsection{Nakahara's Code}
Let $\ell=n=3$.
Then $P_0 = ( 1,0,0)$ and $P_1 = (1/3,1/3,1/3)$ are suitable for $1$-deletion correction.
This code was first found by Dr.\ Mikio Nakahara at Kindai University.
Since $1$-deletion correcting quantum code of length $2$
is prohibited by the quantum no-cloning theorem \cite{wootters82},
this code has the shortest possible length among all $1$-deletion-correcting quantum
codes.

\subsubsection{Example 2}
Let $n=7$, $\ell=3$.
Then $P_0=(7/7,0,0)$, $P_1=(5/7, 1/7,1/7)$, $P_2=(3/7, 2/7,2/7)$
are suitable for $1$-deletion correction.

\subsubsection{Example 3}
Let $n=8$, $\ell=4$.
Then $P_0 = (8/8,0,0,0)$, $P_1=(6/8, 1/8, 1/8,0)$,
$P_2 = (4/8, 4/8, 0, 0)$, $P_3=(4/8, 2/8, 1/8, 1/8)$
are suitable for $1$-deletion correction.

\section{Second Construction of Quantum Deletion Codes}\label{sec4}
The previous construction allows arbitrary $\ell$, but the information rate $(\log_\ell M) / n$
goes to zero as $n\rightarrow \infty$.
In this section, we construct a $t$-deletion-correcting code over $\mathcal{H}_{(t+1)\ell}$,
that is, we assume that the qudit has $(t+1)\ell$ levels.

We introduce an elementary lemma,
which is known in the conventional coding theory \cite{hagiwara20c}.
\begin{lemma}
  Let $\vec{x} = (0$, $1$, \ldots, $t$, $0$, $1$, \ldots $) \in \mathbf{Z}_{t+1}^n$.
  Let $\vec{y}$ be a vector after deletions of  at most $t$ components in $\vec{x}$.
  Then one can determine all the deleted positions from $\vec{y}$.
\end{lemma}
\noindent\textbf{Proof:}
Let $i = \min \{ j \mid y_j > y_{j+1} \}$.
Then $y_1$, \ldots, $y_i$ correspond to $x_1$, \ldots, $x_{t+1}$.
The set difference $\{ x_1$, \ldots, $x_{t+1} \} \setminus \{ y_1$, \ldots, $y_i \}$
reveals the deleted positions among $x_1$, \ldots, $x_{t+1}$.
Repeat the above precedure from $y_{j+1}$ until the rightmost component in $\vec{y}$
and
one gets all the deleted positions.
\rule{1ex}{1ex}

Let $Q \subset \mathcal{H}_\ell^n$ be a $t$-erasure-correcting $((n,M))_\ell$
quantum code.
A codeword $\ket{\psi_1} \in Q$ can be converted to
a codeword in the proposed $t$-deletion-correcting code as follows:
Firstly, observe $\mathcal{H}_{(t+1)\ell}$ is isomorphic to $\mathcal{H}_\ell \otimes
\mathcal{H}_{t+1}$.
Let $\ket{\psi_2} = \ket{01\cdots t 0 1  \cdots } \in \mathcal{H}_{t+1}^{\otimes n}$.
The sender sends $\ket{\psi_1} \otimes \ket{\psi_2}$ as a codeword
in $\mathcal{H}_{(t+1)\ell}^{\otimes n}$.

The receiver receives $\rho \in \mathcal{S}(\mathcal{H}_{(t+1)\ell}^{\otimes n-t'})$,
where $0 \leq t' \leq t$,
where $\mathcal{S}(\mathcal{H}_{(t+1)\ell}^{\otimes n-t'})$ denotes
the set of density matrices on $\mathcal{H}_{(t+1)\ell}^{\otimes n-t'}$.
The quantum system of received state can be divided to
$\mathcal{H}_\ell^{\otimes n-t'}$ and $\mathcal{H}_{t+1}^{\otimes n-t'}$.
The receiver make a projective measurement
on the subsystem $\mathcal{H}_{t+1}^{\otimes n-t'}$
defined by
$\{ \ket{\vec{y}}\bra{\vec{y}}$ $ \mid \vec{y} \in \mathbf{Z}_{t+1}^{n-t'} \}$.
Then the receiver knows all the deleted positions.
After that,
the receiver applies the erasure correction procedure of $Q$,
for example, \cite{matsumoto17uni} for quantum stabilizer codes
  \cite{ashikhmin00,calderbank97,calderbank98,gottesman96,matsumotouematsu00}.

When $\ell$ is a prime power and $t$ is fixed relative to $n$,
$\lim_{n\rightarrow \infty} (\log_\ell M) / n$ can attain $1$ \cite{feng04},
and by the above construction the information rate
$\lim_{n\rightarrow \infty} (\log_{(t+1)\ell} M) / n$ can attain $\log_{(t+1)\ell} \ell$.

\begin{remark}
Let $\rho \in \mathcal{S}(\mathcal{H}_\ell^{\otimes n})$ be a quantum codeword
of an entanglement assisted code \cite{brun06}. By using $\rho$ in place of $\ket{\varphi_1}$
in this section, one can construct $t$-deletion-correcting entanglement assisted code.
\end{remark}

\section{Conclusion}\label{sec5}
This paper proposes two systematic constructions of
quantum deletion-correcting codes.
The first one has advantage of supporting arbitrary dimension of qudits.
The second one has advantages of multiple deletion correction
and asymptotic goodness.
It is a future research agenda to find a construction of having
all the above stated advantages.

\section*{Acknowledgment}
The authors would like to thank Dr.\ Mikio Nakahara at Kindai University
for the helpful discussions.


\begin{thebibliography}{10}
\providecommand{\url}[1]{#1}
\csname url@samestyle\endcsname
\providecommand{\newblock}{\relax}
\providecommand{\bibinfo}[2]{#2}
\providecommand{\BIBentrySTDinterwordspacing}{\spaceskip=0pt\relax}
\providecommand{\BIBentryALTinterwordstretchfactor}{4}
\providecommand{\BIBentryALTinterwordspacing}{\spaceskip=\fontdimen2\font plus
\BIBentryALTinterwordstretchfactor\fontdimen3\font minus
  \fontdimen4\font\relax}
\providecommand{\BIBforeignlanguage}[2]{{%
\expandafter\ifx\csname l@#1\endcsname\relax
\typeout{** WARNING: IEEEtran.bst: No hyphenation pattern has been}%
\typeout{** loaded for the language `#1'. Using the pattern for}%
\typeout{** the default language instead.}%
\else
\language=\csname l@#1\endcsname
\fi
#2}}
\providecommand{\BIBdecl}{\relax}
\BIBdecl

\bibitem{levenshtein66}
V.~I. Levenshtein, ``Binary codes capable of correcting deletions, insertions,
  and reversals,'' \emph{Soviet physics doklady}, vol.~10, pp. 707--710, 1966.

\bibitem{sima21}
J.~Sima and J.~Bruck, ``On optimal $k$-deletion correcting codes,'' \emph{IEEE
  Trans. Inform. Theory}, 2021, doi:10.1109/TIT.2020.3028702, IEEE Early
  Access.

\bibitem{bennett97}
\BIBentryALTinterwordspacing
C.~H. Bennett, D.~P. DiVincenzo, and J.~A. Smolin, ``Capacities of quantum
  erasure channels,'' \emph{Phys. Rev. Lett.}, vol.~78, pp. 3217--3220, Apr.
  1997.
\BIBentrySTDinterwordspacing

\bibitem{grassl97}
\BIBentryALTinterwordspacing
M.~Grassl, T.~Beth, and T.~Pellizzari, ``Codes for the quantum erasure
  channel,'' \emph{Phys. Rev. A}, vol.~56, pp. 33--38, Jul. 1997.
\BIBentrySTDinterwordspacing

\bibitem{macwilliams77}
F.~J. MacWilliams and N.~J.~A. Sloane, \emph{The Theory of Error-Correcting
  Codes}.\hskip 1em plus 0.5em minus 0.4em\relax Amsterdam: Elsevier, 1977.

\bibitem{hagiwara20b}
M.~Hagiwara and A.~Nakayama, ``A four-qubits code that is a quantum deletion
  error-correcting code with the optimal length,'' in \emph{Proc. 2020 IEEE
  International Symposium on Information Theory}, Jul. 2020, pp. 1870--1874.
  DOI: 10.1109/ISIT44484.2020.9174339

\bibitem{hagiwara20a}
A.~Nakamura and M.~Hagiwara, ``The first quantum error-correcting code for
  single deletion errors,'' \emph{IEICE Communications Express}, vol.~9, no.~4,
  pp. 100--104, Apr. 2020.
  DOI: 10.1587/comex.2019XBL0154

\bibitem{hagiwara20c}
A.~Nakayama and M.~Hagiwara, ``Single quantum deletion error-correcting codes,''
  in \emph{Proc. International Symposium on Information Theory and Its
  Applications}, Oct. 2020, pp. 329--333.
  DOI: 10.34385/proc.65.B10-1

\bibitem{hagiwara21}
T.~Shibayama and M.~Hagiwara, ``Permutation-invariant quantum codes for
  deletion errors,'' Feb. 2021, arXiv:2102.03015

\bibitem{picode2}
Y.~Ouyang, ``Permutation-invariant quantum coding for quantum deletion
  channels,'' Feb. 2021, arXiv:2102.02494.

\bibitem{csiszarbook2}
I.~Csisz\'ar and J.~K\"orner, \emph{Information Theory: Coding Theorems for
  Discrete Memoryless Systems}, 2nd~ed.\hskip 1em plus 0.5em minus 0.4em\relax
  Cambridge University Press, 2011.

\bibitem{picode}
Y.~Ouyang, ``Permutation-invariant quantum codes,'' \emph{Phys. Rev. A},
  vol.~90, no.~6, p. 062317, Dec. 2014.

\bibitem{ogawa05}
T.~Ogawa, A.~Sasaki, M.~Iwamoto, and H.~Yamamoto, ``Quantum secret sharing
  schemes and reversibility of quantum operations,'' \emph{Phys. Rev. A},
  vol.~72, no.~3, p. 032318, Sep. 2005.

\bibitem{wootters82}
W.~K. Wootters and W.~H. Zurek, ``A single quantum cannot be cloned,''
  \emph{Nature}, vol. 299, pp. 802--803, 1982.

\bibitem{matsumoto17uni}
R.~Matsumoto, ``Unitary reconstruction of secret for stabilizer based quantum
  secret sharing,'' \emph{Quant. Inf. Process.}, vol.~16, no.~8, p. 202, Aug.
  2017.

\bibitem{ashikhmin00}
A.~Ashikhmin and E.~Knill, ``Nonbinary quantum stabilizer codes,'' \emph{IEEE
  Trans. Inform. Theory}, vol.~47, no.~7, pp. 3065--3072, Nov. 2001.

\bibitem{calderbank97}
A.~R. Calderbank, E.~M. Rains, P.~W. Shor, and N.~J.~A. Sloane, ``Quantum error
  correction and orthogonal geometry,'' \emph{Phys. Rev. Lett.}, vol.~78,
  no.~3, pp. 405--408, Jan. 1997.

\bibitem{calderbank98}
------, ``Quantum error correction via codes over {GF(4)},'' \emph{IEEE Trans.
  Inform. Theory}, vol.~44, no.~4, pp. 1369--1387, Jul. 1998.

\bibitem{gottesman96}
D.~Gottesman, ``Class of quantum error-correcting codes saturating the quantum
  Hamming bound,'' \emph{Phys. Rev. A}, vol.~54, no.~3, pp. 1862--1868, Sep.
  1996.

\bibitem{matsumotouematsu00}
R.~Matsumoto and T.~Uyematsu, ``Constructing quantum error-correcting codes for
  $p^m$-state systems from classical error-correcting codes,'' \emph{IEICE
  Trans.\ Fundamentals}, vol. E83-A, no.~10, pp. 1878--1883, Oct. 2000.

\bibitem{feng04}
K.~Feng and Z.~Ma, ``A finite {Gilbert-Varshamov} bound for pure stabilizer
  quantum codes,'' \emph{IEEE Trans. Inform. Theory}, vol.~50, no.~12, pp.
  3323--3325, Dec. 2004.

\bibitem{brun06}
T.~Brun, I.~Devetak, and M.-H. Hsieh, ``Correcting quantum errors with
  entanglement,'' \emph{Science}, vol. 314, no. 5798, pp. 436--439, 2006.

\end{thebibliography}


\end{document}